\documentclass[aps,prstab,preprint,showpacs,groupedaddress]{revtex4}
\usepackage{graphics,epsfig}

\begin{document}

\title{Dynamics of opinion formation in a small-world network}

\author{Ping-Ping Li$^{1}$, Da-Fang Zheng$^{2}$, and P. M. Hui$^{1}$}
\affiliation{$^{1}$Department of Physics, The Chinese University
of Hong Kong \\
Shatin, New Territories, Hong Kong \\
$^{2}$ Zhejiang Institute of Modern Physics and Department of
Physics, Zhejiang University, HangZhou 310027, People's Republic
of China}

\begin{abstract}
The dynamical process of opinion formation within a model using a
local majority opinion updating rule is studied numerically in
networks with the small-world geometrical property.  The network
is one in which shortcuts are added to randomly chosen pairs of
nodes in an underlying regular lattice.  The presence of a small
number of shortcuts is found to shorten the time to reach a
consensus significantly. The effects of having shortcuts in a
lattice of fixed spatial dimension are shown to be analogous to
that of increasing the spatial dimension in regular lattices.  The
shortening of the consensus time is shown to be related to the
shortening of the mean shortest path as shortcuts are added.
Results can also be translated into that of the dynamics of a
spin system in a small-world network.

\end{abstract}

\maketitle

\section{Introduction}

The physics of networks has received much attention in recent
years.  Topologically, a network consists of nodes and links, with
the latter connecting the nodes in some fashion.  Traditionally,
the network models studied in many branches of science are the
regular lattices and the random networks (classical random graphs)
\cite{erdos}.  It was not until the late 1990's that scientists,
many of them physicists, discovered that many real-world networks
exhibit geometrical properties that are different from regular
networks and random graphs
\cite{review1,review2,mendesbook,wattsbook1,wattsbook2}. Among
these properties are: (i) A small average distance between
arbitrarily chosen nodes in the network, i.e., each node may reach
another node through a path the only passes through a few other
nodes. This is the so-called small-world effect. Typically, the
average distance increases with the number of nodes in the network
only logarithmically.  (ii) The clustering coefficient, which
characterizes the extent in which the connected neighbors of a
node are also connected, in real-world networks is relatively
high.  In the context of a society of individuals, these
properties imply that one can approach any individual through a
few intermediate connections, and the friends of a person are
likely to be friends of each other.

An important branch of research in networks deals with the effects
of the geometrical properties of networks on dynamical processes
in networks.  Taking the nodes as individuals, such dynamical
processes may be epidemics in a population, cultural assimilation,
opinion formation, voting or election, decision making on
competing for limited resources
\cite{pasto,hui1,redner1,redner2,vilone1,vilone2,redner3,hui2,hui3,hui4}.
An up-to-date review on dynamical processes in complex networks
can be found in the recent article by Bocaletti {\em et al.}
\cite{boccaletti}.  An important question is to see how networking
effects may affect the final state of a dynamical process and the
time of reaching the final state.  In the present work, we study
the dynamical process of opinion formation based on a recent model
of Redner and coworkers \cite{redner1,redner2} in which the
dynamics is based on a local majority rule \cite{galam}.  Although
the model was original stated in terms of spins (nodes) that can
take on one of two possible states, we will instead describe the
model in terms of opinion formation between agents (nodes) who can
take on one of two opinions.  The translation from one description
to another is obvious.  The local majority rule then describes the
influence of neighboring (connected) agents on an agent's opinion.
In particular, we will study the changes in the dynamical process
as a regular lattice is transformed into a small-world network by
adding links to connect randomly chosen pairs of nodes
\cite{newman}.  This model of small-world network is analogous to
that of Watts and Strogatz \cite{WS}, and this underlying network
has been used to investigate voting processes
\cite{vilone1,vilone2} and the Ising model \cite{boyer}.  We found
that the time to reach a consensus drops sensitively with the
additional of a small fraction of links to an otherwise regular
lattice.  The additional of links in a lattice of fixed spatial
dimension is found to have similar effects as increasing the
spatial dimensionality of a regular lattice and to bring the
system closer to the mean field results previously obtained in the
literature \cite{redner1,redner2}.  It should be noted that the
model studied here is based on a local majority rule, which is
different from another popular model of opinion formation called
the voter model \cite{vilone1,redner3}.  The latter has the
advantage of being analytically manageable, while analytic
treatments of the local majority model on complex networks remains
a challenging task.  While previous work on the voter model
\cite{redner3} focused on the effects of the spread in degrees in
complex networks on the consensus time, here we found that there
exists strong correlation between the change in the time to reach
a consensus and the mean shortest distance of the underlying
network.

The plan of the paper is as follows.  We introduce the model of
opinion formation and the underlying network structure in Sec.II.
Results of detailed simulations are presented and discussed in
Sec.III.  A summary of the results is given in Sec.IV.

\section{Opinion formation model and Network Structure}

We consider a system with $N$ nodes, which may, for example,
represent $N$ agents or $N$ magnetic moments or spins depending on
the situation under consideration.  For each node, there are two
possible states which are represented by $+1$ and $-1$. These
states represent two opposite opinions.  Following the model
studied by Redner and co-workers \cite{redner1,redner2}, the
states of the nodes evolve in time according to the following
updating rules.  At each time step, one node is chosen randomly.
The chosen node and his connected neighbors through an underlying
network are then considered collectively for updating.  For
hypercubic lattices in $D$-dimension, for example, the cluster
size is $(2D+1)$ nodes since each node is connected to $2D$
nearest neighboring nodes. All the nodes in the cluster of nodes
will then be updated to take on the state of the local majority.
The updating rule thus represents a consensus is reached in a
cluster by taking the majority opinion. The procedure is then
repeated until all the nodes reach a common state, i.e., when a
final state of consensus is reached.

Previous studies \cite{redner1,redner2} on the model were carried
out on regular lattices.  Recent research on the science of
networks reveals that many real-world networks exhibit the
small-world effect.  The effect refers to the shortest distance
from one node to another in networks.  In contrast, the distance
in a regular lattice grows with the size of the lattice. To study
the effects of shortening of distances on opinion formation, we
choose a lattice proposed by Newman and Watts \cite{newman} in
which one may study the effects of a gradual change in the
distance as links are randomly added to an underlying regular
lattice.  We start with a two-dimensional (2D) square lattice of
size $\sqrt{N} \times \sqrt{N}$ with periodic boundary conditions.
A number of additional links, called ``shortcuts", are added
between a randomly chosen pair of nodes.  Slight modifications to
the model in Ref.\cite{newman} are that the shortcuts cannot link
a node to itself and at most one link is allowed between any two
nodes, i.e., no doubly connected nodes. To quantify the number of
additional links on the lattice, we introduce a parameter $q$,
which is the number of additional links normalized by the total
number of links $2N$ in the underlying lattice. For given $q$, the
number of additional links is $2qN$.  Note that in changing a
square lattice into a {\em fully connected} network without doubly
connected nodes, a total of $N(N-5)/2$ additional links are
needed.  For given $q$, the corresponding fraction of all possible
additional links is, therefore, given by $4q/(N-5)$. From our
numerical results, we found that additional links amount to $q
\approx 1$ are sufficient for studying the networking effects in
the dynamics of opinion formation.  Thus, we will focus in the
range $0 \leq q \lesssim 1$ in the following discussions.  To
allow for comparison with results on regular lattices, we randomly
choose $2D$ neighbors among the neighbors of the chosen site for
updating in each time step.  We have checked that using the whole
cluster of a chosen site for updating give nearly identical
results.

\section{Results}
\subsection{Shortening of consensus time}
An important quantity for opinion formation is the consensus time,
which is the time for the system to reach a common opinion.
Starting with an initial configuration of a fraction $p$ of $+1$
and a fraction of $(1-p)$ of $-1$ states among the nodes, the time
to reach consensus is studied numerically.  Figure 1(a) gives the
results of the mean consensus time $T$ in a 2D lattice with
different numbers of shortcuts in units of Monte Carlo step per
site, i.e., one such time step corresponds to the time during
which each node would have been updated once on average. Results
are obtained by averaging 5000 runs with different initial
configurations for given values of $p$ and $q$. The $q=0$ results
corresponding to that on a square lattice.  The results show that
as a small number ($q < 1$) of shortcuts are added, the consensus
time drops sensitivity for nearly the whole range of $p$. For a
small range near $p \approx 0$ and $p \approx 1$, additional links
may lead to a slightly longer consensus time, but just by a tiny
bit, as a result of possible re-enforcement of the survival of
some scattered minority states by the additional links.  In our
model, we have chosen to fix the updating cluster size to be
$(2D+1)$, where $D$ is the dimension of the underlying regular
lattice. It will be interesting to check the results against those
in which the chosen node is updated together with {\em all} its
connected nodes. Figure 1(b) shows the results for such a model,
with the normalization to the simulation time steps taken to be
$(\langle k \rangle + 1)$, where $\langle k \rangle$ is the mean
degree of the network after the shortcuts are added. The results
of the two models are nearly identical. The reason is that the
degree distribution of the network with added links, unlike the
scale-free networks \cite{barabasi}, has a sharp and
characteristic peak. The drop in the consensus time is most
sensitive in the intermediate range of $p$. For $p \approx 0.5$, a
drop of consensus time of two orders of magnitude can be achieved
by the addition of about $N$ links to the system ($q=0.5$) (see
Fig.1(a)).  Interestingly, recent studies on the dynamics of voter
models \cite{vilone1,vilone2} in 1D small-world networks revealed
that ordering processes become harder to achieve with the addition
of shortcuts.  The ordering processes in Ising model was also
found to be slower \cite{boyer} in small-world networks.  For the
majority updating rule studied here, the shortcuts {\em
accelerates} the ordering process of reaching a common opinion.

The sensitivity of the consensus time to $q$ may be a result of a
qualitative change in the time dependence of the dynamics. We
studied in detail how the number of nodes with $+1$ state changes
as a function of time for various values of $p$. Figure 2 shows
the results for $p=0.3$ in both 2D and 1D networks, which are
typical of the intermediate range of $p$.  For $p=0.3$ in a 2D
lattice with $N=2500$ nodes, the final consensus state has all the
nodes with $-1$ state. The fraction $n_{+}$ of nodes taking on
$+1$ state therefore drops as a function of time.  For finite $q$,
the drop in $n_{+}$ is much more rapid than $q=0$ and takes on an
exponential decaying behavior.  We have checked that similar
$q$-dependence of $n_{+}(t)$ results in underlying lattice of
higher spatial dimensions.  It is useful to show the corresponding
result in 1D (Fig.2(b)), as results for $n_{+}(t)$ have also been
reported in 1D networks for the voter model \cite{vilone1}.
Comparing the results in 1D and 2D networks for the majority rule
model, the behavior is qualitatively the same, but the time it
takes for $n_{+}(t)$ to vanish is much longer in 1D.  In general,
the addition of shortcuts shortens the consensus time
significantly. Comparing with results in the voter model
\cite{vilone1}, there are similarities as well as qualitative
differences.  For both models, the addition of shortcuts leads to
a shorter consensus time when compared with a regular lattice
($q=0$).  However, the two models are {\em qualitatively
different} in the details of the dynamics. For the model based on
local majority rule (see Fig.2), $n_{+}(t)$ drops more rapidly
with time for $q\neq 0$ for {\em all time} $t$ during the time
evolution, when compared with that in a regular lattice ($q=0$).
For the voter model, there exists a significant duration of time
(in particular early stage) during the evolution of $n_{+}(t)$
that $n_{+}(t)$ drops {\em slower} in a network with shortcuts
($q\neq 0$) than in a regular lattice ($q=0$) -- a result clearly
shown in Fig.3 of Ref.\cite{vilone1}. It is only at the later
stage of the time evolution of $n_{+}(t)$ that $n_{+}(t)$ drops
rapidly and the consensus time eventually becomes shorter than
that in a regular lattice.  While it is tempting to draw close
analogies between the two models, it is also important to note
that they are essentially different models and they show different
dynamics in the process of evolving to a consensus.

\subsection{Shortcuts and mean field limit}

To further explore the effects of the shortcuts, we recall that in
Ref.\cite{redner1} a mean field limit of the opinion formation
model was studied.  The mean field limit corresponds to a model in
which at each time step, a group of $G$ nodes are chosen {\em at
random} among all the $N$ nodes in the network for updating. The
underlying network can thus be thought of one in which every node
is possibly connected to another.  In Ref.\cite{redner1}, the
authors compared simulations results in 1D, 2D, 3D, and 4D
lattices without shortcuts and found that the behavior tends to
approach the mean field limit as dimensionality increases. They
found that the mean field behavior has not been reached in 4D and
thus a higher upper critical dimension is expected for the
problem.  Figure 3 shows the exit probability, which is the
probability that the final state of the system is one with all the
nodes taking on $+1$ state, as a function of $p$ for 1D and 2D
underlying lattices with added shortcuts. The inset shows the
results of 2D networks.  The $q=1$ results overlap with the mean
field results, which are obtained by numerical simulations by
randomly choosing groups of $G=5$ nodes for updating in every time
step. Analytically, the exit probability takes on the form of an
error function \cite{redner1}.  Our results, therefore, indicate
that the effects of $q$ in a lattice with {\em fixed} spatial
dimension are similar to that of increasing dimensionality in
regular lattices \cite{newman}, i.e., increasing $q$ has the
effect of bringing the system closer to the mean field limit.  To
explore this effect more carefully, Fig.3 (main panel) shows the
results in a 1D network with $100$ nodes for different values of
$q$, together with results of the mean field limit.  The updating
group size is $3$ for each time step. It is clear from the results
that as $q$ increases, the exit probability changes more abruptly
near $p=0.5$ and the results approach the mean field results. The
relation between an increasing $q$ and an effective dimension has
been studied by Newman and Watts \cite{newman}. It was pointed out
that there exists a characteristic length scale $\xi$ in the
network model which gives the typical distance between the ends of
the shortcuts on the lattice.  The length $\xi$ plays the role of
the correlation length in problems related to phase transitions
and critical phenomena.  The effective dimension can be extracted
by imposing the relation $N \sim L^{D}$, where $N$ is the number
of nodes in the network and $L$ is the mean shortest path between
two nodes. For an underlying regular lattice in $d$-dimension of
linear size $S$, the effective dimension $D$ is related to $d$
through \cite{newman} $D = d \log(S/\xi)$ for $\xi \ll S$, and
$D=d$ for $\xi \gg S$.  Since $\xi \sim (qd)^{-1/d}$, $D = d$ only
for a very dilute fraction of shortcuts, i.e., in the $q
\rightarrow 0$ limit.  For a large range of $q$, we are in the
regime of $\xi \ll S$ and hence the effective dimension $D$ is
higher than the dimension of the underlying lattice $d$.

The consensus time varies in different realizations corresponding
to given values of $q$ and $p$.  The consensus time distribution
for the $q=0$ (no added links) case consists of a mean peak and a
long tail. The latter corresponds to runs in which clusters of
minority states are formed so that a longer time is needed to
change the minority opinion of the clusters. We define the {\em
most probable} consensus time $T_{mp}$ as the time corresponding
to the peak of the distribution \cite{redner1}. For different
values of $q$, we studied the distribution and obtained $T_{mp}$
for networks of different sizes $N$. It is also observed that for
$q$ slightly larger than zero, the tail in the consensus time
distribution shrinks rapidly, and hence the mean consensus time
and $T_{mp}$ become closer in value for finite $q$. Figure 4 shows
the results of $T_{mp}$ as a function of $N$ for different values
of $q$ in a log-log plot, together with the results in the mean
field limit \cite{redner1}. The $q=0$ results reproduce the
behavior reported in Ref.\cite{redner1}, with $T_{mp} \sim
N^{\alpha}$ and $\alpha = 1.11 \pm 0.01$. As $q$ increases, the
values of $\alpha$ decreases.  The trend of decreasing $\alpha$
has been reported for regular lattices as dimensionality increases
\cite{redner1}. It was reported the mean field limit has not been
reached in 4D regular lattices.  In contrast, we found that even a
small value of $q =1$ leads to a value of $\alpha$ quite close to
the mean field limit.  The observation is again related to the
increasing effective dimension as $q$ increases, as discussed
above.

\subsection{Consensus time and shortest distance}

The mean field limit corresponds to networks with easy access from
one node to another.  The effects of adding links to a regular
lattice lead to a similar effect in that the distances between
nodes drop rapidly with the number of added links \cite{newman},
thus leading to the small-world effect.  This effect is
illustrated in Fig.5(a) by showing the drop in $L' = L/ln N$ as a
function of $q$ for lattices of $N=3600$ and $N=10000$ nodes,
where $L$ is the mean shortest distance of the network obtained by
averaging the shortest distances between all pairs of nodes in the
network.  Besides a $ln N$ dependence, $L$ drops with $q$.  It is
then interesting to correlate the mean consensus time and the
shortest distance for a given value of $q$.  In Fig.5(b), the
dependence of the mean consensus time $T' = T/(Nln N)$ on $q$ is
shown for two values of $N$, for the case of $p=0.5$.  Dividing
the mean consensus time $T$ by $N ln N$ removes the $N$ dependence
and only $q$ dependence remains. Note that both $L'$ and $T'$ show
similar dependence on $q$. To illustrate that the shortening of
the mean consensus time is intrinsically a result of the
shortening of the shortest distance when shortcuts are added, we
plot $T'$ as a function of $L'$ in Fig.5(c). The data for small
$L'$ correspond to that of $q \gtrsim 0.6$.  We see that $T'$
increases monotonically with $L'$.

\section{summary}
We have studied the effects of adding shortcuts connecting
randomly chosen pairs of sites in a regular lattice on the
consensus time in reaching a common opinion within a model of
local majority updating rule.  The consensus time is found to drop
sensitively with the addition of a small number of shortcuts. This
drop is observed nearly over the whole range of $p$ characterizing
the initial distribution of opinions among the nodes.  The rapid
drop of consensus time with the addition of shortcuts is shown to
be related to a qualitative change for finite $q$ in the time
dependence of the drop of the number of minority nodes as a
function of time, as compared with regular lattice without
shortcuts.  We also compared our results with mean field results
previously reported in the literature.  It is found that the
addition of shortcuts in a lattice of fixed spatial dimension has
the similar effects of increasing the spatial dimension of regular
lattices.  This similarity is shown in both the behavior of the
exit probability and the dependence of the most probable consensus
time on the network size.  Geometrically, the shortcuts decreases
the mean shortest distance in a network and leads to the
small-world effect.  Dynamically, these shortcuts bring in the
opinion of more connected nodes to a chosen node and hence have
the effects of dissolving local minority groups more rapidly.  The
consensus time is found to change monotonically with the shortest
distance in the network.

\acknowledgments{The work was completed during a recent visit of
one of us (D.F.Z.) to the Chinese University of Hong Kong under
the support of a Direct Grant of Research at CUHK. P.M.H.
acknowledges the partial support of a Grant from the Research
Grants Council of the Hong Kong SAR Government under  grant number
CUHK-401005. D.F.Z. also acknowledges the partial support of the
National Natural Science Foundation of China under Grant nos.
70471081, 70371069 and 10325520.}

\newpage
\centerline {FIGURE CAPTIONS}

Figure 1: Consensus time $T$ in units of Monte Carlo step per site
as a function of the initial fraction $p$ of $+1$ states in a
$N=50\times50$ square lattice for $q=0, 0.1, 0.5$, and $1$,
respectively (from top to bottom).  (a) The updating group size is
fixed to be $2D+1=5$ with the chosen node plus $2D$ randomly
selected neighbors of the chosen node. (b) Updating process
involves the chosen node and all its connected nodes.

Figure 2: The fraction $n_{+}$ of nodes taking on $+1$ state as a
function of time $t$ (in units of Monte Carlo step per site) for
$p=0.3$ and $q=1$, $0.5$, $0.1$, and $0$ (from left to right) in
(a) a $N=50\times50$ square lattice and (b) a $N=100$ 1D lattice.

Figure 3: Exit probability $E$ as a function of $p$ for
one-dimensional (main panel) and two-dimensional networks (inset)
for different values of $q$.  The 1D networks are of size $N=100$
nodes and the 2D networks are of size $N=50 \times 50$. The
symbols correspond to $q=0$ (squares), $0.1$ (circles), $0.5$
(triangles), $1$ (inverted triangles).  The mean field results
(diamonds) refer to those obtained by randomly choosing groups of
$G=2D+1$ nodes for updating in each time step.

Figure 4: Most probable consensus time $T_{mp}$ in units of Monte
Carlo step per site as a function of network size $N$ for the
initial state corresponding to $p=0.5$.  The data from top to
bottom correspond to $q=0$, $0.05$, $0.1$, $0.5$ ,$1$, and the
mean-field limit, respectively. The lines represent linear fits
with slopes $1.11\pm0.01$, $0.61\pm0.02$, $0.45\pm0.02$,
$0.20\pm0.01$, $0.167\pm0.007$, and $0.158\pm0.007$.

Figure 5: (a) The scaled mean shortest distance scaled $L^{\prime}
= L/(\ln N)$ as a function of $q$, for two-dimensional networks
with $N=3600$(squares) and $N=10000$ (circle) nodes. (b) The
scaled mean consensus time $T^{\prime}=T/(N \ln N)$ in units of
Monte Carlo step per site as a function of $q$ for initial state
with $p=0.5$ for two-dimensional networks of sizes $N=3600$
(squares) and $N=10000$ (circles) nodes. (c) $T^{\prime}$ depends
monotonically on $L^{\prime}$.

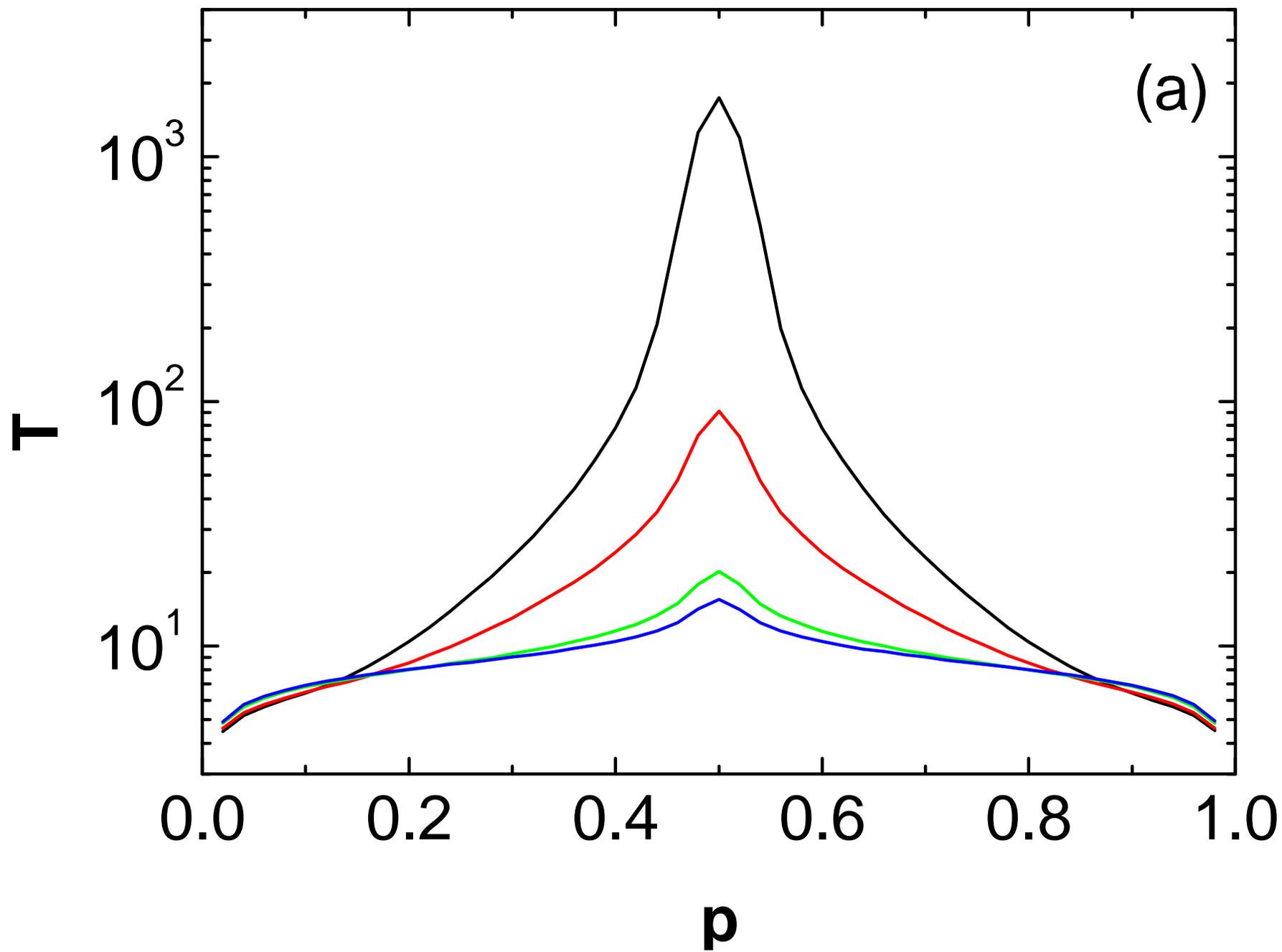
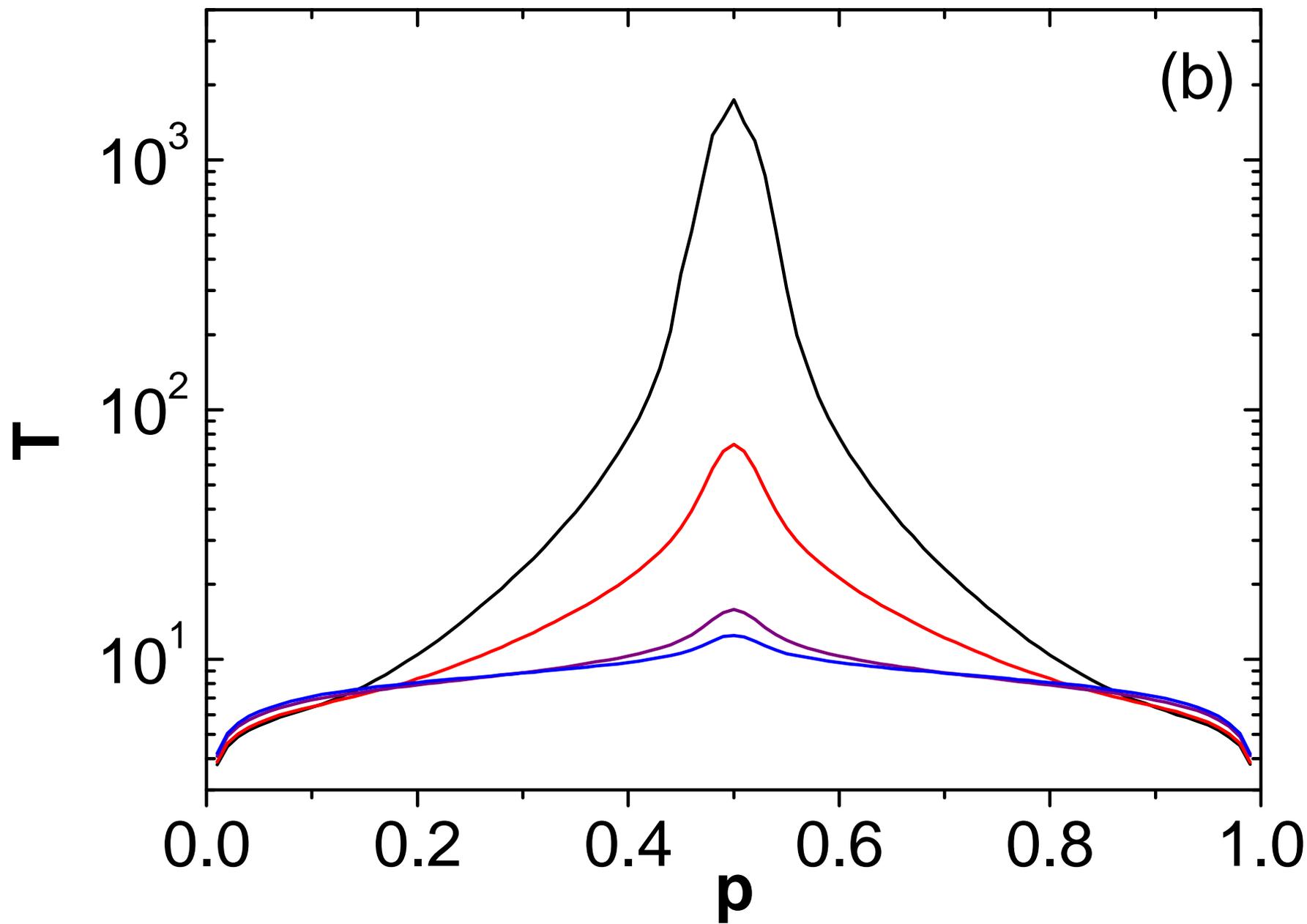
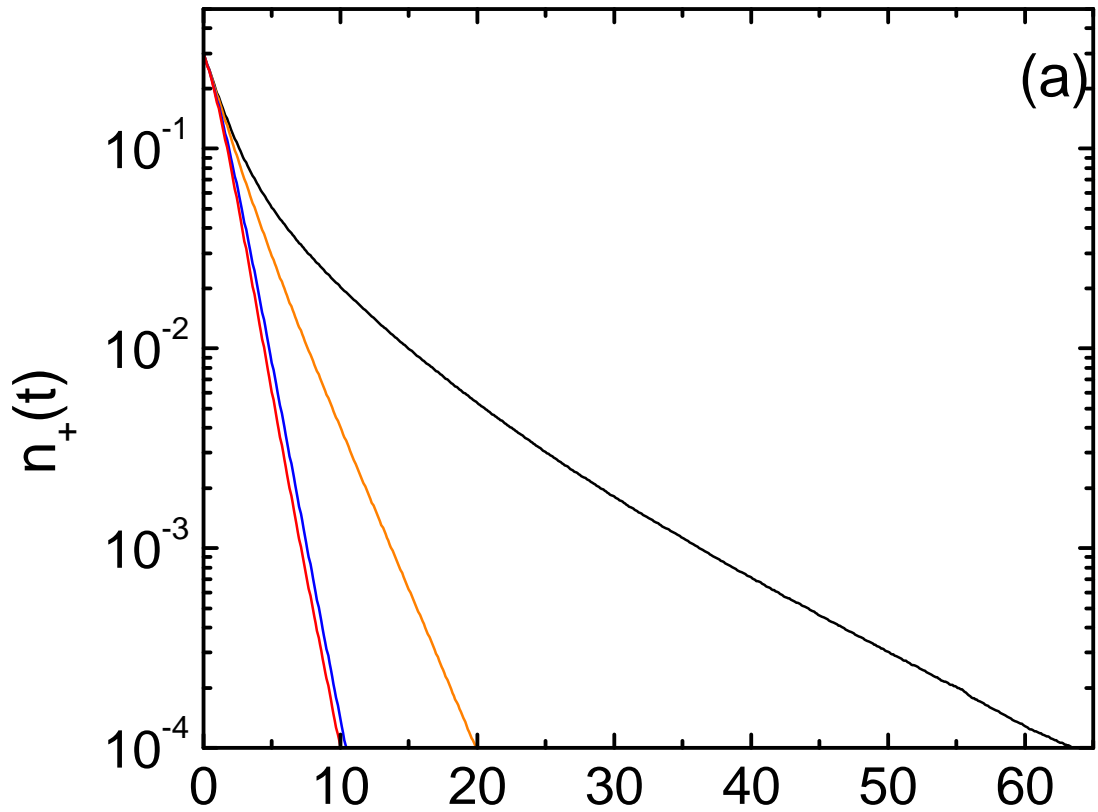
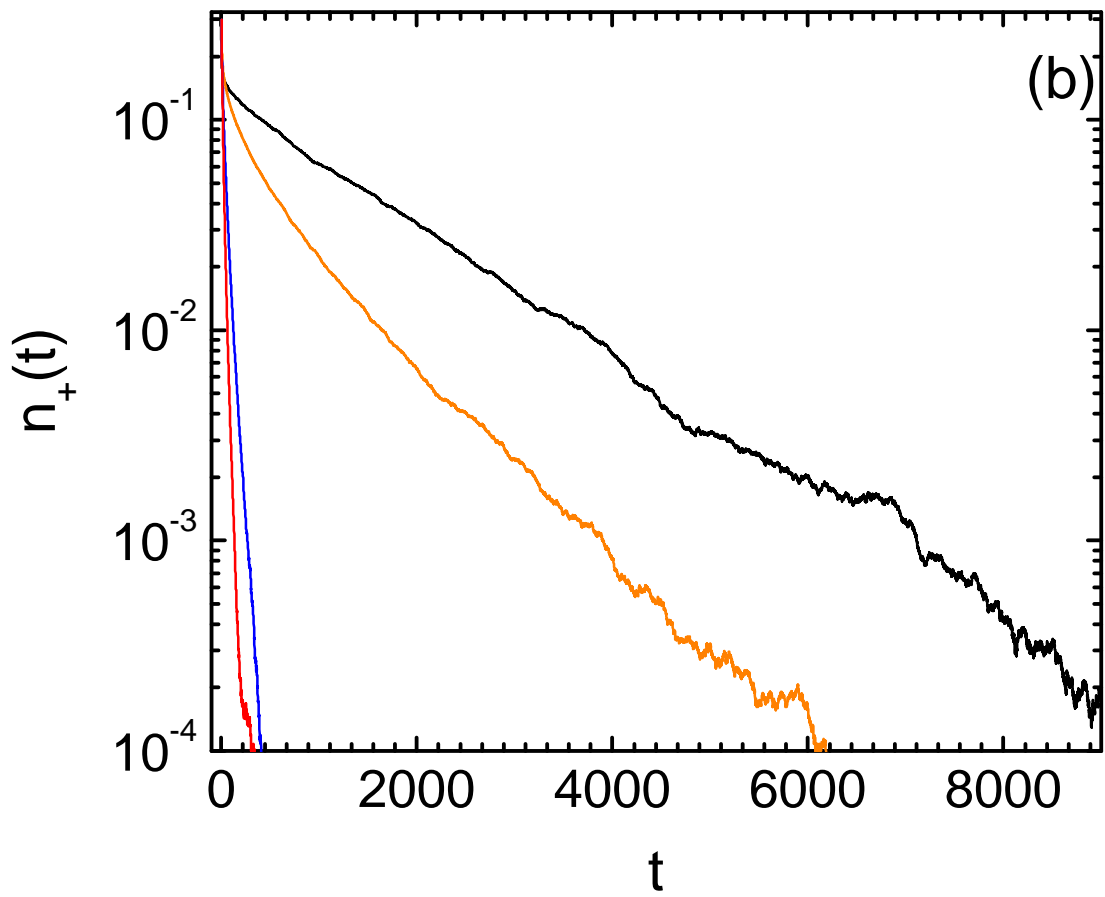
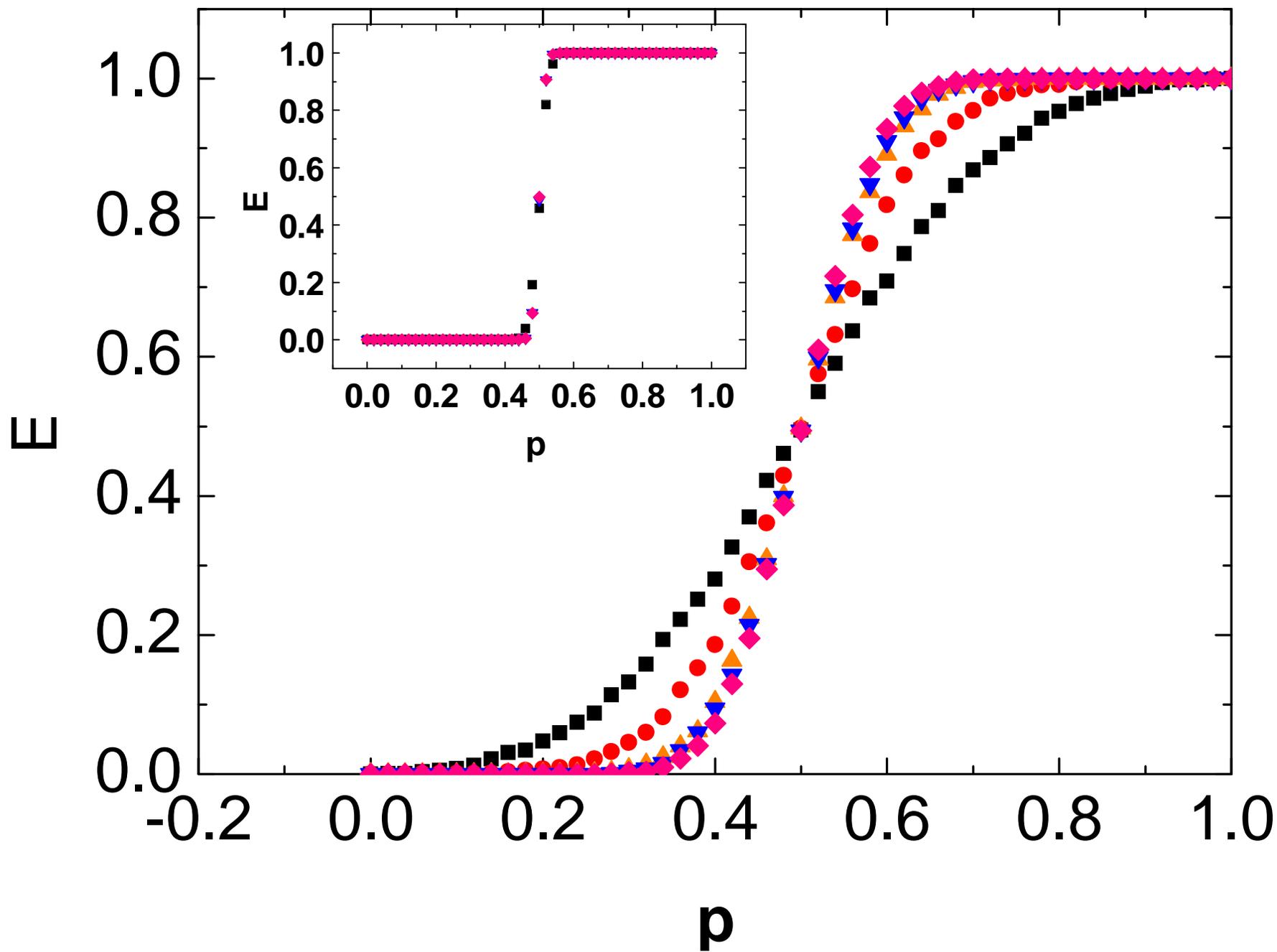
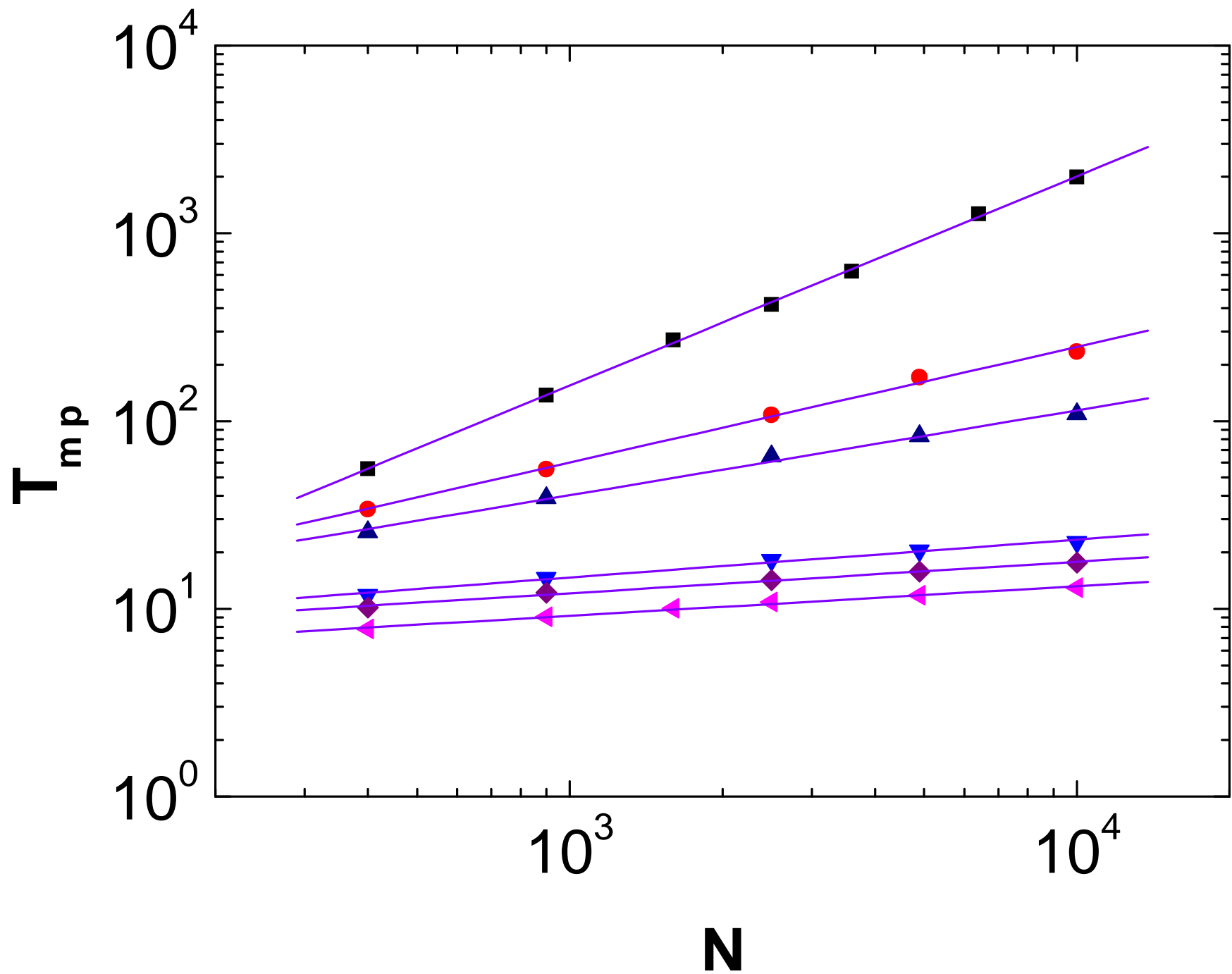
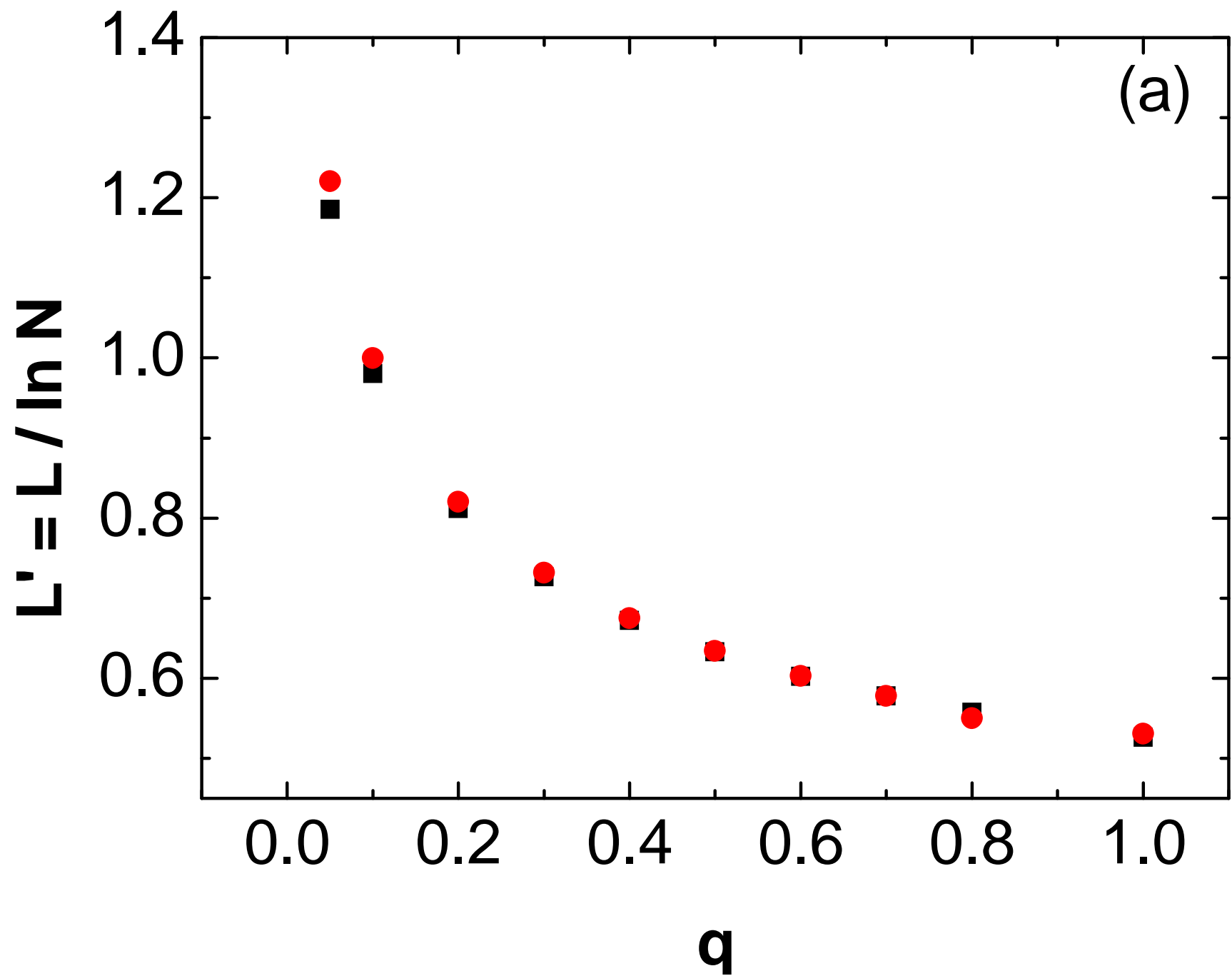
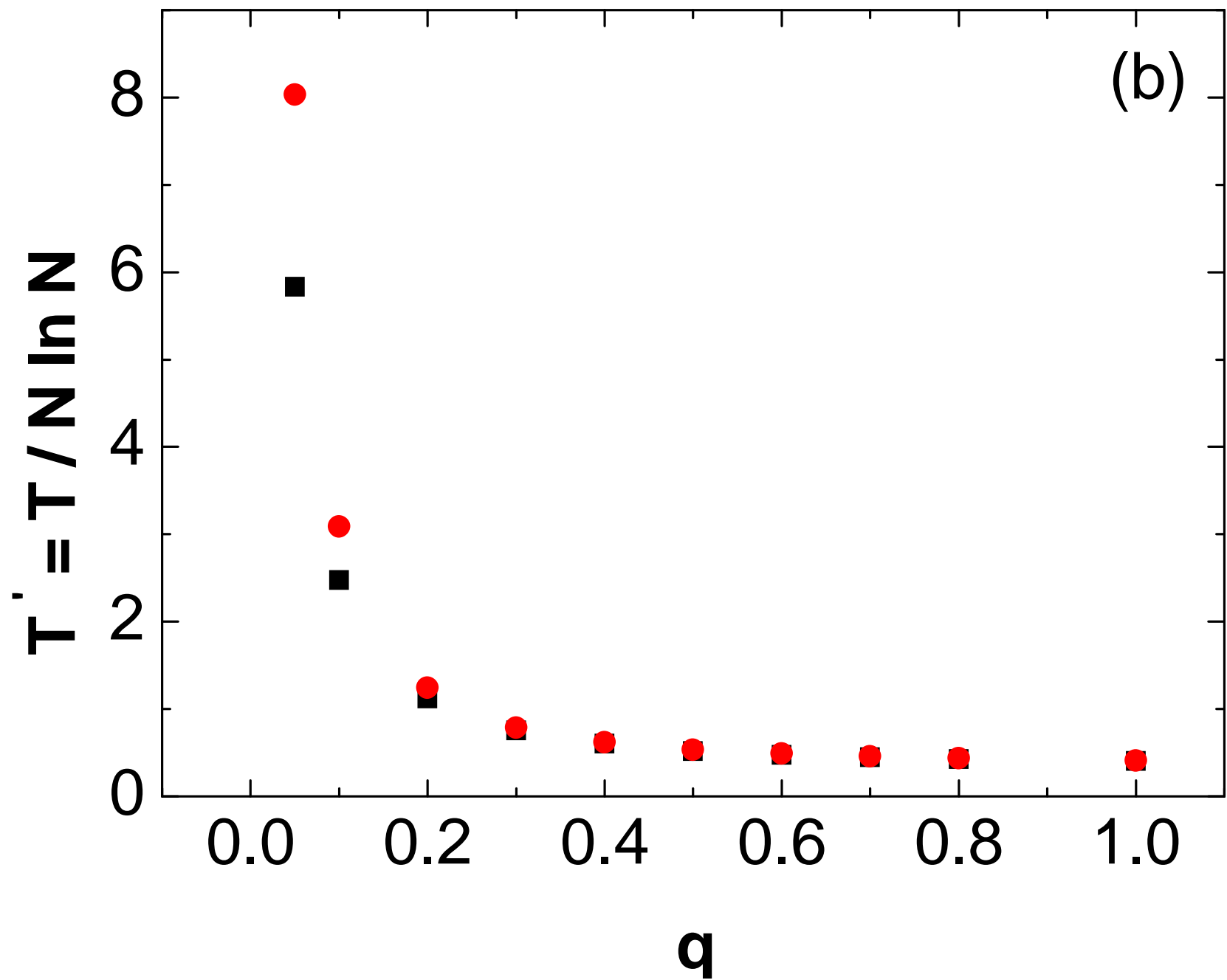
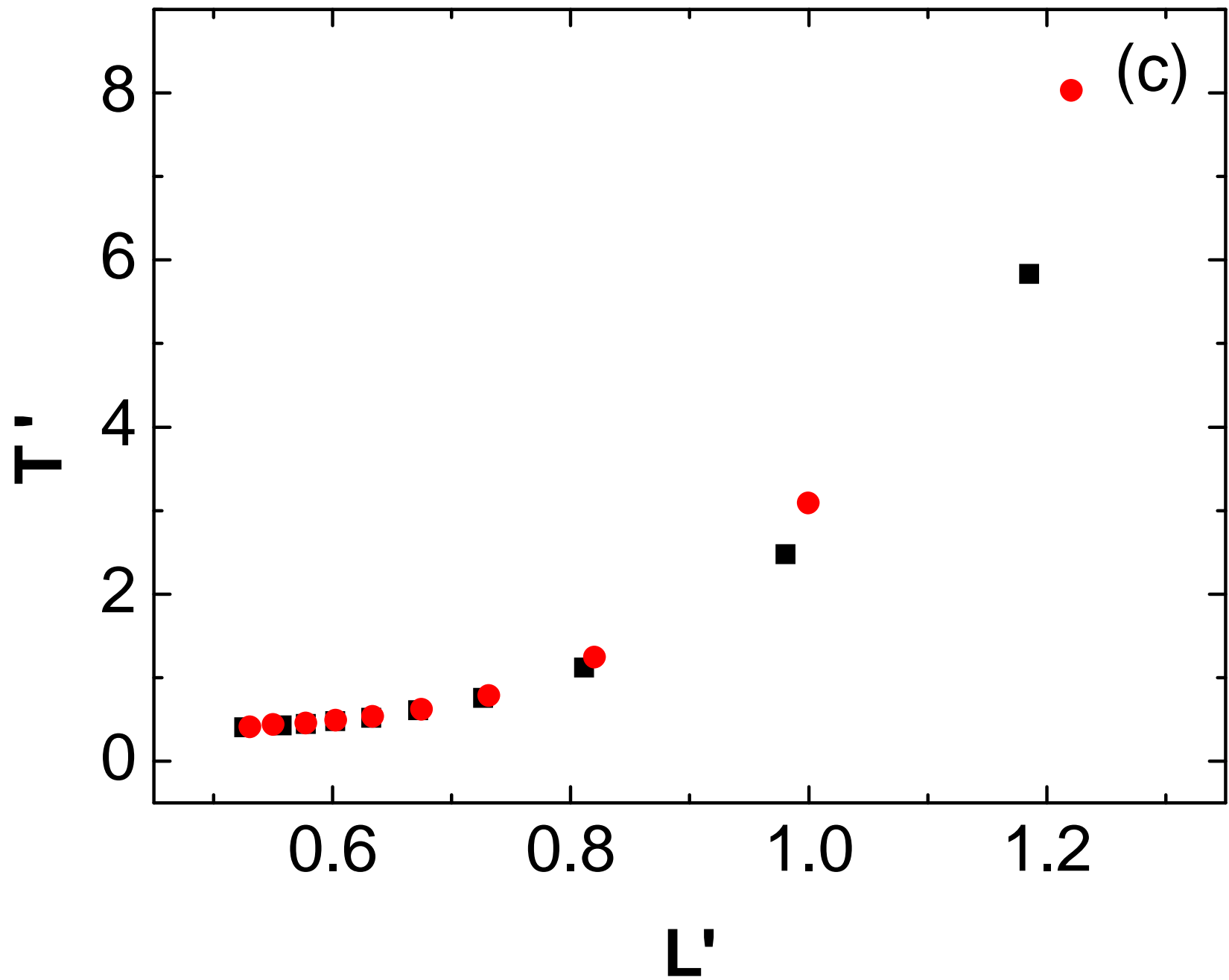
\end{document}